\documentclass[aps,prl,reprint,groupedaddress]{revtex4-1}
\usepackage{amsmath,amssymb,bm}
\usepackage{graphicx}

\pdfoutput=1
\usepackage{braket}
\usepackage{amsmath}

\begin{document}

\title{Magnetic susceptibility of quantum spin systems calculated by sine square deformation: 
one-dimensional, square lattice, and kagome lattice Heisenberg antiferromagnets} 

\author{Chisa Hotta}
\affiliation{Department of Basic Science, University of Tokyo, 3-8-1 Komaba, Meguro, Tokyo 153-8902, Japan}
\email{chisa@phys.c.u-tokyo.ac.jp}
\author{Kenichi Asano}
\affiliation{Center for Education in Liberal Arts and Sciences, Osaka University, Toyonaka, Osaka 560-0043, Japan}
\date{\today}

\begin{abstract}
We develop a simple and unbiased numerical method to obtain the uniform susceptibility 
of quantum many body systems. 
When a Hamiltonian is spatially deformed by multiplying it with {\it a sine square function} 
that smoothly decreases from the system center toward the edges, 
the size-scaling law of the excitation energy is drastically transformed to a rapidly converging one. 
Then, the local magnetization at the system center becomes nearly size independent; 
the one obtained for the deformed Hamiltonian of a system length as small as $L\sim 10$ 
provides the value obtained for the original uniform Hamiltonian of $L\sim 100$. 
This allows us to evaluate a bulk magnetic susceptibility by using the magnetization at the center 
by existing numerical solvers without any approximation, parameter tuning, or the size-scaling analysis. 
We demonstrate that the susceptibilities of the spin-1/2 antiferromagnetic Heisenberg chain 
and square lattice obtained by our scheme at $L\sim 10$ 
agree within $10^{-3}$ with exact analytical and numerical solutions for 
$L=\infty$ down to temperature of $0.1$ times the coupling constant. 
We apply this method to the spin-1/2 kagome lattice Heisenberg antiferromagnet 
which is of prime interest in the search of spin liquids. 
\end{abstract}





%
\maketitle
{\em Introduction.---}
Computing the thermodynamic properties of a many-body quantum lattice model 
over a wide range of temperature is a challenging problem, which remains too often unsolved. 
Prominent examples include quantum spin systems with nontrivial ground states 
such as spin liquids\cite{sl}, as found in the kagome lattice antiferromagnet\cite{kagome} and Kitaev model\cite{kitaev}. 
Experimentally, much effort has been devoted to measuring the magnetic susceptibility of relevant materials such as 
ZnCu$_3$(OH)$_6$Cl$_2$\cite{helton07,mendels10},  BaCu$_3$V$_2$O$_8$(OH)$_2$ \cite{hiroi09}, 
and $\kappa$-ET$_2X$\cite{shimizu03,yyoshida15}, to get the smoking guns of their realization. 
In real materials, the interesting physics is always found at temperatures ($T$) much lower than the 
characteristic interaction, $J$; 
indeed, the $T$ dependence of susceptibility contains rich information such as 
whether or not the excitations are gapped, 
spinons or majorana fermions form a Dirac point or a Fermi surface. 
However, numerical methods such as exact diagonalization(ED)\cite{ftlczs} 
and typicality approaches\cite{hans93,ebi03,shimizu13} 
suffer from severe finite size effects and cannot capture the behavior at $T<J$. 
The quantum Monte-Carlo(QMC) method gives reliable results down to $T \sim 0.1J$ \cite{okabe88,cyasuda05}, 
but is not applied to most of the above mentioned nontrivial models because of the sign problem. 
\par
The high-temperature series expansion(HTE)\cite{htebook} serves as a powerful analytical tool complementary to numerics.
However, the series in powers of $\beta=1/k_BT$ extends at most up to $\beta^{16}$ to $\beta^{19}$\cite{oitmaa96,elstner94}, 
and falls off from the true result at $T\lesssim J$. 
To further extend a series down to $T\sim J/2$, 
a numerical linked cluster (NLC) approach has been considered\cite{rigol06}, 
and the entropy method\cite{bernu15} succeeded in interpolating 
between the $T=0$ limit and a HTE result for $T>J$. 
However, these methods are still delicate at present since they are based on some assumptions. 
For instance, the entropy method requires {\it a priori} knowledge of 
the susceptibility near $T=0$ based on the ground state information. 
It is highly desirable to have {\it vice versa}, 
i.e., to extract such low $T$ information from the thermodynamic observables. 
\par
Given such situation, a reliable and practical approach that is valid at any temperature is desperately needed. 
Here we propose {\it a parameter-free} and {\it unbiased} scheme to obtain susceptibility 
by making use of a device called sine square deformation (SSD)\cite{nishino09}. 
The SSD is a spatial modification of the energy scale of the Hamiltonian. 
It serves as one of the boundary conditions\cite{gendiar11,hikihara11,katsura11,okunishi15}, 
as well as works as a real-space renormalization scheme\cite{ch2}. 
It also reveals itself as one of a low energy effective Hamiltonian in a 2D conformal field theory
\cite{belavin,katsura12,okunishi16}. 
Moreover, adiabatic connections between the uniform and SSD Hamiltonian is guaranteed\cite{tada}. 
\par 
{\em SSD.---} We first introduce the SSD Hamiltonian, 
in which an envelope function $f_{\rm SSD}$ makes the original Hamiltonian 
${\cal H}=\sum_i (h(\bm r_i)-\mu n(\bm r_i))$ spatially nonuniform: 
\begin{eqnarray}
&&{\cal H}_{\rm SSD}= \sum_{\bm i}f_{\rm SSD} (\bm r_i)\big( h(\bm r_i)-\mu n(\bm r_i) \big), 
\label{hssd}
\\
&& f_{\rm SSD} (\bm r_i)= \frac{1}{2}\Big(1+\cos \big(\frac{\pi r_i}{R}\big)\Big). 
\label{fssd}
\end{eqnarray} 
Here, $\bm {r}_i$ a coordinate of the lattice site if $h(\bm r_i)$ is an on-site term, 
and it is a coordinate of the bond if $h(\bm r_i)$ is an inter-site interaction or a hopping term. 
The origin of $\bm r_i$ is at the center of the cluster\cite{ssd-radial}. 
In $f_{\rm SSD}(\bm r_i)$, $R$ is chosen to be slightly larger than $R_0$, the distance from 
the system center to the farthest edge site. 
If the Hamiltonian is written in terms of fermionic operators, $\mu$ is the chemical potential, 
and $n(\bm r_i)$ is a particle density. 
If the Hamiltonian is writtten in terms of spin operators, then $\mu$ 
and $n(\bm r_i)$ are replaced with magnetization $m(\bm r_i)$ and magnetic field $H$, respectively. 
We solve ${\cal H}_{\rm SSD}$ and evaluate the expectation values of local quantities $A(\bm r_i)$ 
for energy eigenstates, which are no longer translationally invariant. 
\par
One of our previous findings was that the ground state physical quantities evaluated at the system center, 
where $f_{\rm SSD} (\bm r_i) \sim 1$, are nearly independent of the system size $N$ 
and mimic the values for $N\rightarrow\infty$ for the original Hamiltonian\cite{ch1}. 
For example, by applying a magnetic field, $H$, to quantum magnets, 
one can compute a magnetization density $\langle m (\bm r_i=0) \rangle$ 
for the SSD ground state\cite{ch1}. 
Even for system lengths $L \lesssim 20$ ($N=L^d$ for $d$ dimension), 
we obtain a magnetization curve mimicking the bulk exact solution of the original uniform Hamiltonian 
within an accuracy of 10$^{-4}$ in 1D\cite{ch1}, and $10^{-3}$ in 2D\cite{ch2,ch3}. 
\par
Intuitively, deforming a Hamiltonian may mean modifying the physical system itself, but for SSD, 
this is not the case\cite{ch2,maruyama11}. 
We have shown earlier that the modified part of the Hamiltonian, ${\cal H}_{\rm SSD}-{\cal H}$, 
renormalizes the energy levels of the original ${\cal H}$ in a way similar to the poor man's scaling by Wilson\cite{ch2}. 
The excitation energy, $\epsilon_l(L)$, follows a $1/L^2$ behavior\cite{ch2,tamura17} 
and densely populates around $\epsilon=0$, 
in sharp contrast to the standard scaling law, $1/L$. 
As a result, by using a system size as small as $L\sim 10$ in SSD system, 
one can suppress finite size effects down to those of the original Hamiltonian for $L\sim 100$. 
%
\begin{figure}[tb]
\centering
\includegraphics[width=0.49\textwidth]{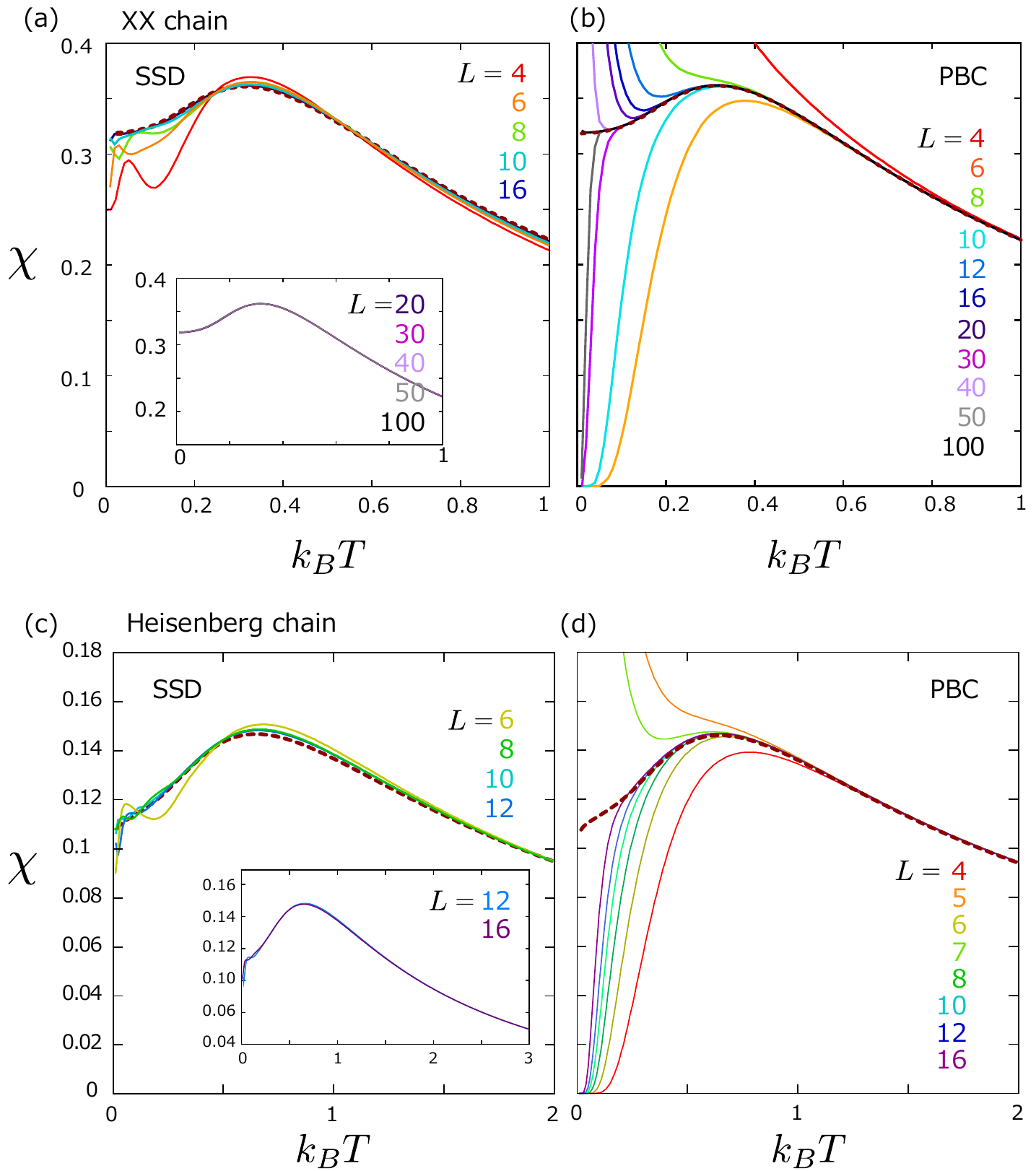}
\caption{
Susceptibility $\chi$ of the 1D XX model, which is equivalent to a free fermionic chain, 
obtained by (a) our scheme using SSD and (b) the standard method 
in a uniform system with a periodic boundary condition (PBC). 
$\chi$ of the spin-1/2 Heisenberg chain, obtained by (c) our scheme 
and (d) the standard method with a PBC.  
Broken line is the exact analytical solution for $L=\infty$\cite{klumper}. 
All the results are numerically exact. 
}
\label{f1}
\end{figure}
\par
However, it is not clear whether the whole excited state spectrum is well preserved by SSD. 
Here we show that is the case. 
As a result, thermodynamic quantities are very accurately calculated, practically free of size effects. 
\par
{\em The local Gibbs ensemble.---} 
Consider a lattice consisting of $N=L^d$ sites and deform a Hamiltonian following Eq.(\ref{hssd}). 
By solving ${\cal H}_{\rm SSD}$ at finite temperature, 
obtain the Gibbs ensemble, $\langle \cdots \rangle$, of a {\it local physical quantity defined 
at the system center} with index $c$ as 
\begin{equation}
\langle \hat A_c \rangle = \frac{1}{\Xi} \sum_l \langle \psi_l | \hat A_c |\psi_l \rangle {\rm e}^{-\beta E_l} ,
\label{ai}
\end{equation}
where $\Xi=\sum_{l} {\rm e}^{-\beta E_l}$ is the grand partition function, 
and $\phi_l$ the many body wave function with energy $E_l$. 
Our main conclusion is that, for fermionic systems, 
the particle density at the center, $\langle \hat n_c \rangle$, 
for system sized as small as $N\gtrsim 10$ in 1D and $N \gtrsim 20$ in 2D 
agree with those for the original Hamiltonians for $N \sim \infty$ within $\sim 10^{-3}$. 
This conclusion holds also for the magnetization, $\langle \hat m_c \rangle$, of spin systems. 
Once energy and particle density, or magnetization, are obtained as 
{\it smooth functions of $\beta$ and $\mu$, or $H$}, 
thermodynamic potentials and {\it all} thermodynamic quantities can be evaluated. 
\par
{\em Noninteracting system.---} Let us first demonstrate the validity of our claim 
for the quantum $S=1/2$ XX spin chain, 
${\cal H}=\sum_i ( s_i^x s_{i+1}^x+s_i^y s_{i+1}^y - H s_i^z )$, 
which is equivalent to a free fermionic chain. 
By applying a small magnetic field, $H=0.01\sim 0.1$, 
we obtain an exact solution of the SSD Hamiltonian for a given $L$, 
evaluate $\langle \hat m_c \rangle$, and take its derivative to obtain a uniform static susceptibility, 
$\chi=d\langle \hat m_c \rangle/dH$. 
Figure~\ref{f1}(a) shows the result from $L=6$ up to 100. 
Already at $L\sim 8$, they are in good agreement with the exact $L=\infty$ susceptibility of the bulk XX spin chain in broken line.
Remarkably, the gapless behavior, $\chi >0$ at $T\rightarrow 0$, is correctly obtained 
even for $L=4$. 
When $L\sim 10$ the accuracy of $\chi$ already reaches $10^{-3}$ at $k_BT \sim 0.2$ and $10^{-4}$ at higher temperatures. 
\par
By contrast, $\chi$ of the original Hamiltonian obtained in the standard manner, 
$\chi=-\beta^{-1} \langle (\sum_i m_i)^2 \rangle^2/N$ (with $H=0$), suffers from a serious finite-size effect (Fig.\ref{f1}(b)). 
As a consequence of energy gap in the low energy spectrum, $\Delta_L \sim {\cal O}(1/L)$, 
$\chi$ shows an artificial exponential drop, $\propto {\rm e}^{-\beta \Delta_L}$, at $k_BT \sim 0$. 
\par
{\em Heisenberg systems.---} Our scheme yields similarly high accuracies for interacting systems. 
For the spin-1/2 Heisenberg chain, ${\cal H}=\sum_{i} s_i s_{i+1} - H \sum_i s^z_i$, 
we perform a full ED for $L\le 16$  
and adopt a typicality approach called the thermal pure quantum (TPQ) method for $L>16$\cite{shimizu13}\cite{tpq}, 
to solve ${\cal H}_{\rm SSD}$. 
Figures~\ref{f1}(c) shows $\chi$ obtained with our scheme. 
Direct comparison with the the exact solution for $L=\infty$\cite{klumper} 
shows that our results are accurate within the order of $10^{-4}$ for $L>10$. 
Moreover, even for $L$ as small as 8, 
a small drop of $\chi$ appears at temperatures lower than $k_B T \sim 0.01$, reminiscent of 
the well-known logarithmic singularity in Bethe ansatz solution\cite{log}. 
For the original Hamiltonian, 
finite-size effects again lead to an artificial exponential drop, as shown in Fig.~\ref{f1}(d). 
\par
\begin{figure}[tb]
\centering
\includegraphics[width=0.49\textwidth]{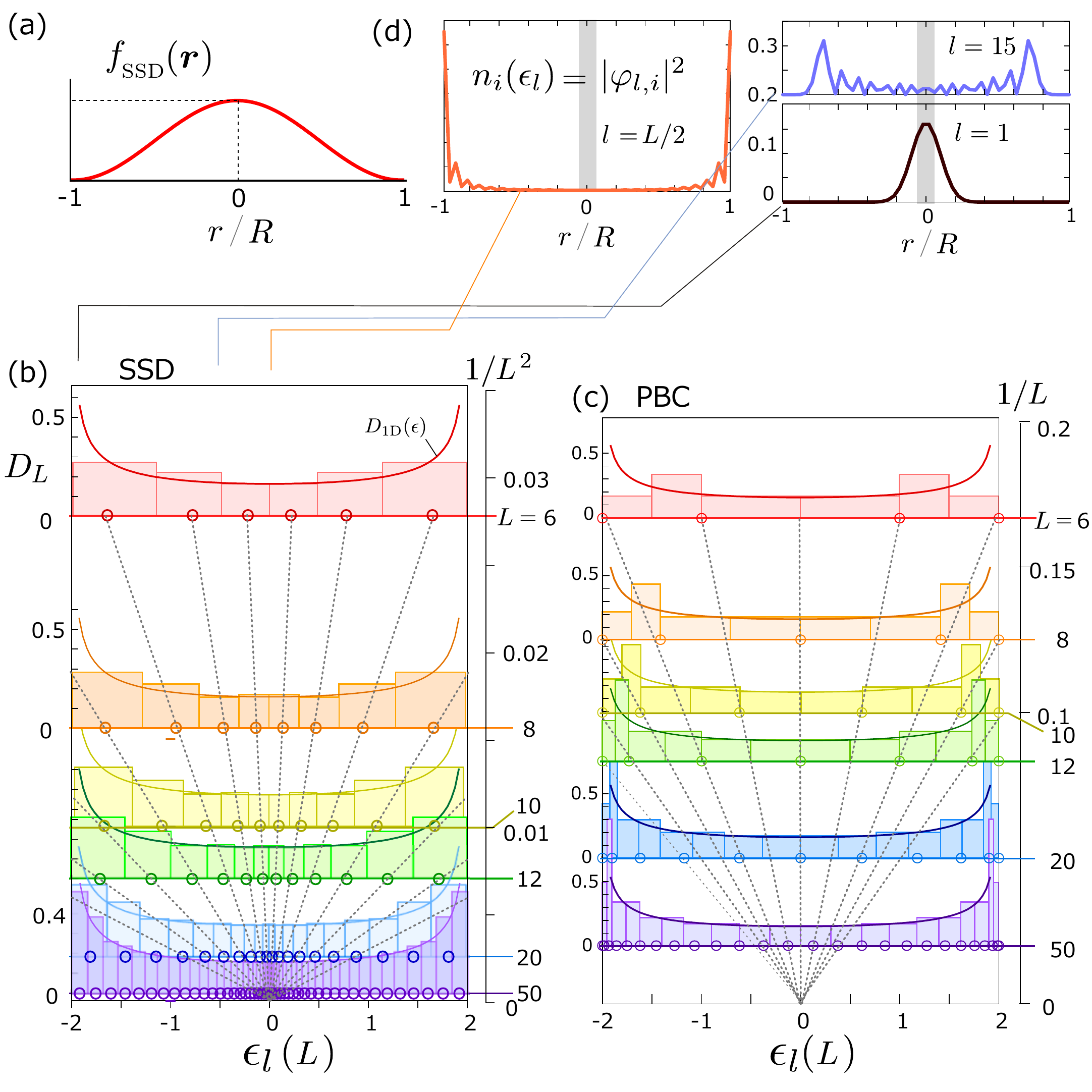}
\caption{
(a) SSD function, $f_{\rm SSD}(\bm r)$ in 1D. 
In (b) and (c), eigenenergies $\epsilon_l$ (horizontal axes) of the SSD and original (PBC) Hamiltonians 
of free fermionic chain are shown as circles for several choices of $L$, as a function of 
$L^{-2}$ and $L^{-1}$ (vertical axes), respectively. 
Colored vertical bars are the effective one-body DOS of 1D free fermions, $D_L(\epsilon_l)$; 
in (b) they include information on the particle density at the center of the SSD system, 
and in (c) they are the usual discrete finite size DOS. 
Colored solid lines in (b) and (c) are the bulk exact density of states of 1D free fermions, $D_{1D}(\epsilon)$.
(d) Spatial distribution of the particle density of the SSD free fermionic chain of $L=50$
for the $l=$25, 10, and 1, with $l=1$ being the lowest energy. 
}
\label{f2}
\end{figure}
\begin{figure}[tb]
\centering
\includegraphics[width=7.5cm]{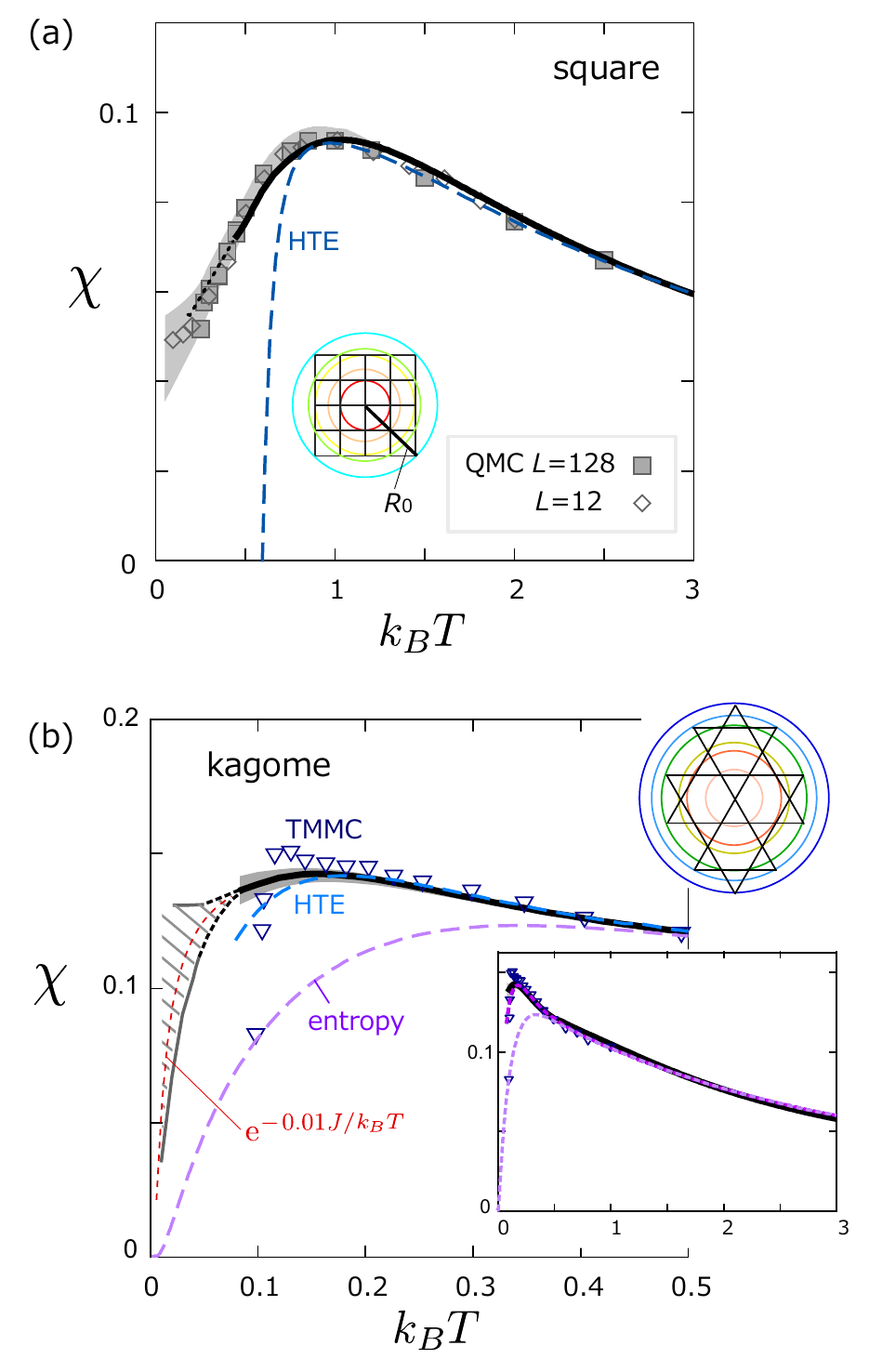}
\caption{
Susceptibility, $\chi$, of spin-1/2 2D Heisenberg models on (a) square and (b) kagome lattices 
of finite sizes and shapes shown in the insets, calculated with SSD, 
where colored circles are the guide to the eye to clarify which sites and bonds belong 
to the same radius $\bm r_i$ in $f_{\rm SSD}(\rm r_i)$. 
In each panel, solid black line is our result, obtained for an optimal 
$dR$ and making averages of $\sim 20$ TPQ samples for (a), and $\sim 60$ TPQ samples for (b). 
Shading at low $T$ indicates uncertainties due to large spatial oscillations of $\langle m(\bm r)\rangle$ near $\bm r=0$ 
that varies from TPQ sample to sample. 
The hatched region in panel (b) indicates the same ambiguity 
in our ED results, for which 100 lowest states were used. 
Symbols and broken lines indicate previous works:  
[(a) QMC\cite{wang92,okabe88}, HTE\cite{sqhte}] and 
[(b) TMMC\cite{tota92}, entropy\cite{bernu15}, HTE\cite{lohmann14}]. 
The lower inset to (b) shows our, as well as earlier, results to higher temperatures.
}
\label{f3}
\end{figure}
\par
{\em Density of states.---} We now clarify how the ensemble in Eq.(\ref{ai}) works 
for a 1D free fermionic chain. 
Let us deform the Hamiltonian 
${\cal H}=\sum_i (-c_i^\dagger c_{i+1} + {\rm h.c.} - \mu n_i)$, 
and then diagonalize it into, 
${\cal H}_{\rm SSD}=\sum_l (\epsilon_l-\mu) a^\dagger_l a_l$ 
where the creation operator $a_l^\dagger$ is related to $c_i^\dagger$ by the unitary transformation 
$a_l^\dagger= \sum_i \varphi_{l,i} c_i^\dagger$. 
The distribution of the one-body eigenenergy, $\epsilon_l(L)$, ($l=1$ through $L$), 
for system length $L$ is shown in Fig.\ref{f2}(b). 
One finds a clear $L^{-2}$ dependence (broken lines) starting from $\mu$ ($\epsilon_l(L)=0$). 
This is in sharp contrast to the $L^{-1}$ behavior for the original Hamiltonian 
shown in Fig. \ref{f2}(c) 
which is known from the conformal field theory for 1D quantum critical systems\cite{cardy}. 
\par
The $l$-th one-body eigenstate has a particle density at the $i$-th site given by 
$n_i(\epsilon_l)\equiv |\varphi_{l,i}|^2$. 
In Fig.\ref{f2}(d) we show the $i$-dependence of $n_i(\epsilon_l)$ 
for three different energy levels $\epsilon_l$ for $L=50$:
the chemical potential level ($l=L/2$), 
a slightly lower level($l=15$), and the band bottom($l=1$). 
For the original Hamiltonian the particle density does not depend on $\bm r_i$. 
However, SSD gives each location $i$ its own energy scale proportional to $f_{\rm SSD}(\bm r_i)$. 
Consequently, the particles distribute in a way forming a wave packet, which has large weight at $\bm r_i$ that 
overall fulfills the relation $f_{\rm SSD}(\bm r_i)\sim |\epsilon_l-\mu|$; 
at $\epsilon_l \sim \mu$, 
the wave function forms an edge state in which $f_{\rm SSD}(L/2)\sim 0$, 
whereas near the band edge with maximum $|\epsilon_l|$, the particle is localized at the center. 
Since $f_{\rm SSD} \sim 1$ at the system center (the shaded region in Fig.\ref{f2}(d)), 
it is naturally expected that $n_c$ is roughly the same as that for the original Hamiltonian. 
We therefore define an effective density of states (DOS) for system size $L$ as, 
$D_L(\epsilon_l)= \big(\sum_{j=l-1,l,l+1}n_c(\epsilon_{j}) \big)/(\epsilon_{l+1}-\epsilon_{l-1})$. 
Figure~\ref{f2}(b) shows $D_L(\epsilon)$ 
as vertical bars for each $L$, demonstrating that it agrees well with the exact 1D DOS, 
$D_{\rm 1D}(\epsilon)=(2\pi)^{-1} (1-(\epsilon+\mu)^2/4)$, 
shown as solid lines, even for $L$ as small as 4. 
\par
For the original Hamiltonian, $D_L(\epsilon)$ gives the conventional discrete DOS, 
since $n_c$ is a constant filling factor of fermions. 
As shown in Fig.\ref{f2}(c), it is in good agreement with $D_{\rm 1D}(\epsilon)$ for $L>10$ except for a finite size effect. 
However, there is a distinct difference from SSD one in how it is constructed; 
For the original Hamiltonian, the particle density is uniform and the DOS is simply an inverse of 
the discrete energy level spacings. 
By contrast, the SSD compresses the spacings between low energy levels 
and at the same time redistributes the particle density at the system center 
in a way that depends on their energy. These two effects result in a DOS that well reproduces $D_{1D}(\epsilon)$. 
As mentioned earlier, the SSD makes $\epsilon_l(L)$ proportional to $L^{-2}$, 
leading to rapid increase of the number of energy levels in the vicinity of the chemical potential. 
\par
These finding indicate that the particle density of the original Hamiltonian for $L\rightarrow \infty$  
can be obtained from the fictitious DOS represented by $n_c(\epsilon_l)$, 
equivalent to $D_L(\epsilon_l)$, 
\begin{equation}
\langle n (\mu,\beta)\rangle = \sum_{l=1}^L f(\epsilon_{l})  n_c(\epsilon_l) ,
\label{freef}
\end{equation}
where $f(\epsilon)=({\rm e}^{-\beta (\epsilon-\mu) } +1)^{-1}$ is the Fermi distribution function. 
Let us consider an alternative expression in the many body form. 
Construct a many body wave function, $|\psi_l(n_f)\rangle$, ($l=1$ through $\Lambda_{n_f}$), 
with eigen energy $E_l$, where $\Lambda_{n_f}$ is the size of the Hilbert space for a given particle number $n_f$. 
One can easily confirm that 
\begin{equation}
\langle n (\mu,\beta)\rangle = \frac{1}{\Xi_L} \sum_{n_f=0}^{L} \sum_{l=1}^{\Lambda_{n_f}} 
\langle \psi_l(n_f)| \hat n_c |\psi_l(n_f)\rangle {\rm e}^{-\beta E_l}, 
\label{ncmb}
\end{equation}
with $\Xi_N= \sum_{n_f,l} {\rm e}^{-\beta E_l}$, 
gives exactly the same result as Eq.(\ref{freef}). 
This formula is equivalent to Eq.(\ref{ai}), with $\hat A_c=\hat n_c$. 
\par
{\em 2D Heisenberg systems.---}
We apply our scheme to 2D systems. 
Figure~\ref{f3}(a) shows $\chi$ of the spin-1/2 square lattice Heisenberg antiferromagnet, 
in which we have chosen a system size of $N=5\times 5=25$ lattice. 
Here, since TPQ method allows only a very small $L$ in 2D, 
physical quantities oscillate as a function of $\bm r_i$ at $k_BT \lesssim 0.5$ due to boundary effects. 
As already known from the grand canonical analysis at $T=0$, 
the center of oscillation is the true result we need to obtain\cite{ch1}. 
The amplitude of oscillation depends on the choice of $R$. 
We tune the radius $R=R_0+dR$ with $dR$ ranging from 0 to 1 
to minimize such oscillations\cite{supple}. For each $R$, we make a 25 sample average 
of initial TPQ states. 
The shaded region indicates the uncertainty due to large oscillations inevitable at low $T$. 
For comparison, we also plot previous results of QMC calculations for $N=128 \times 128$\cite{ding92} and 
$12\times 12$\cite{okabe88}, and HTE\cite{wang92,sqhte}, 
which is in agreement with our results typically within ${\cal O}(10^{-3})$\cite{accuracy}. 
Here, the HTE provides a very useful check of the accuracy of ourresult for $k_BT \gtrsim 2$. 
\par 
We finally present our unbiased susceptibility of 
spin-1/2 antiferromagnetic Heisenberg kagome lattice in Fig.~\ref{f3}(b). 
It is widely studied by a transfer-matrix Monte Carlo(TMMC)\cite{tota92}, 
NLC\cite{rigol07}, HTE\cite{lohmann14}, and entropy methods assuming gapless excitations\cite{bernu15}. 
Our result supports the strong enhancement at $k_BT\lesssim 0.5$, which is clearly seen in the inset. 
Previous results except the one from the entropy method show similar enhancement. 
Moreover, as shown in the main panel, our $\chi$ starts to drop at $k_BT\sim 0.1$, 
and is slightly smaller than the TMMC result. 
In this region, HTE is no longer reliable. 
At $k_BT<0.1$ we have separately performed ED on ${\cal H}_{\rm SSD}$ for the lowset 100 states 
and evaluated the range of $\chi$, indicated by hatching, 
in order to clarify whether the spin gap is finite or not. 
The range of hatching indicates the ambiguity arising from large oscillation of $\langle m(\bm r_i)\rangle$ at 
$r_i \sim 0$, which increases at lower $T$. 
We plot $\chi \sim {\rm e}^{-\Delta/k_BT}$, which is expected for the spin gapped system, 
and find that even if the gap were finite, it should be as small as $\Delta/J \sim 0.01-0.02$. 
\par
There are many ED studies on 2D quantum magnets that calculate 
$\chi$ and specific heat at $k_B T \lesssim 0.1$. 
However, at such low temperatures, finite size effects become a serious problem. 
Our nearly size-dependence-free scheme also suffers from this limitation. 
Since our scheme is compatible with any numerical solver, 
it should be able to attack this extremely difficult temperature region once 
a powerful solver is developed that can handle twice as large a system than currently possible. 

\begin{acknowledgments}
We thank Johannes Richter and Laula Messio for useful discussions and 
Yasu Takano for helping us with many advices. 
This work was supported by JSPS KAKENHI Grants JP16K05425, JP17K05533, JP17K05497, and JP17H02916. 
\end{acknowledgments}


\begin{thebibliography}{99}
\bibitem{tpq}TPQ method \cite{hans93,ebi03,shimizu13} 
indicates a method that evaluates the finite temperature physical properties 
by a few numbers of states with a reasonably small error. 
By repeatedly operating the Hamiltonian to initial random vector, one can generate a series of pures states 
that represent thermal equilibrium states at various temperatures. 
The evaluation of physical quntities are made by the canonical ensemble of these pure states. 
We prepared more than 20 initial random vectors, and made iteration over 300 times to reach $k_BT\sim 0.01$.
\bibitem{ssd-radial}In two-dimension, we take $f_{\rm SSD}(\bm r_i)$ a radial function instead of the product of sine functions 
in two different directions, since it is shown to be more efficient at $T=0$ in Ref.[\cite{ch3}].
\bibitem{accuracy} In the grand canonical analysis in Ref.[25,30] using SSD, 
the best accuracy of quantities of the 2D ground state was ${\cal O}(10^{-3})$, 
which also holds for the present scheme. This accuracy allows us to evaluate most of the physical properties numerically.
\end{thebibliography}
\end{document}